\documentclass[a4paper,preprint,showpacs,amssymb,pre,superscriptaddress]
              {revtex4}
\usepackage{graphicx}

\begin{document}

\title{System Size Stochastic Resonance: General Nonequilibrium
Potential Framework}

\author{B. von Haeften}
\affiliation{Departamento de F\'{\i}sica, FCEyN, Universidad
Nacional de Mar del Plata\\ De\'an Funes 3350, 7600 Mar del Plata,
Argentine}
\author{G. Iz\'us\cite{conicet}}\email{izus@mdp.edu.ar}
\affiliation{Departamento de F\'{\i}sica, FCEyN, Universidad
Nacional de Mar del Plata\\ De\'an Funes 3350, 7600 Mar del Plata,
Argentine} \affiliation{IMEDEA (CSIC - UIB), E-07122 Palma de
Mallorca, Spain}
\author{H. S. Wio\cite{conicet}}
\email{wio@ifca.unican.es} \affiliation{Instituto de F\'{\i}sica
de Cantabria, Universidad de Cantabria \\ E-39005 Santander,
Spain} \affiliation{Centro At\'omico Bariloche and Instituto
Balseiro, 8400 San Carlos de Bariloche, Argentine}

\begin{abstract}
We study the phenomenon of system size stochastic resonance within
the nonequilibrium potential's framework. We analyze three different
cases of spatially extended systems, exploiting the knowledge of
their nonequilibrium potential, showing that through the analysis of
that potential we can obtain a clear physical interpretation of this
phenomenon in wide classes of extended systems. Depending on the
characteristics of the system, the phenomenon results to be
associated to a breaking of the symmetry of the nonequilibrium
potential or to a deepening of the potential minima yielding an
effective scaling of the noise intensity with the system size.
\end{abstract}


\maketitle

\section{Introduction}

The phenomenon of \textit{stochastic resonance} (SR) ---namely,
the \textit{enhancement} of the output signal-to-noise ratio (SNR)
caused by injection of an optimal amount of noise into a nonlinear
system--- configures a counterintuitive cooperative effect arising
from the interplay between \textit{deterministic} and
\textit{random} dynamics in a \textit{nonlinear} system.  The
broad range of phenomena for which this mechanism can offer an
explanation can be appreciated in  Ref.\cite{RMP} and references
therein, where we can scan the state of the art.

Most of the phenomena that could possibly be described within a SR
framework occur in \textit{extended} systems: for example, diverse
experiments were carried out to explore the role of SR in sensory
and other biological functions \cite{biol} or in chemical systems
\cite{sch}. These were, together with the possible technological
applications, the main motivation to many recent studies showing
the possibility of achieving an enhancement of the system response
by means of the coupling of several units in what conforms an
\textit{extended medium}
\cite{extend1,otros,extend2,extend2b,extend3a,extend3b}, or
analyzing the possibility of making the system response less
dependent on a fine tuning of the noise intensity, as well as
different ways to control the phenomenon \cite{claudio,nos3}.

In previous papers
\cite{extend2,extend2b,extend3a,extend3b,extend3c} we have studied
the stochastic resonant phenomenon in extended systems for the
transition between two different patterns, exploiting the concept
of \textit{nonequilibrium potential} (NEP) \cite{GR,I0}. This
potential is a special Lyapunov functional of the associated
deterministic system which for nonequilibrium systems plays a role
similar to that played by a thermodynamic potential in equilibrium
thermodynamics \cite{GR}. Such a nonequilibrium potential, closely
related to the solution of the time independent Fokker-Planck
equation of the system, characterizes the global properties of the
dynamics: that is attractors, relative (or nonlinear) stability of
these attractors, height of the barriers separating attraction
basins, and in addition it allows us to evaluate the transition
rates among the different attractors \cite{GR,I0}. In another
recent paper we have explored the characteristics of this SR
phenomenon in an extended system composed by an ensemble of
noise-induced nonlinear oscillators coupled by a nonhomogeneous,
density dependent diffusion, externally forced and perturbed by a
multiplicative noise, that shows an effective noise induced
bistable dynamics \cite{WeNew}. The stochastic resonance between
the attractors of the noise-induced dynamics was theoretically
investigated in terms of a two-state approximation. It was shown
that the knowledge of the exact NEP allowed us to completely
analyzed the behavior of the output SNR.

Recent studies on biological models of the Hodgkin-Huxley type
\cite{SSSR1,SSSR2} have shown that ion concentrations along
biological cell membranes present intrinsic SR-like phenomena as
the number of ion channels is varied. A related result
\cite{SSSR3} shows that even in the absence of external forcing,
the regularity of the collective firing of a set of coupled
excitable FitzHugh-Nagumo units results optimal for a given value
of the number of elements. From a physical system point of view,
the same phenomenon --that has been called \textit{system size
stochastic resonance} (SSSR)-- has also been found in an Ising
model as well as in a set of globally coupled units described by a
$\phi ^{4}$ theory \cite{SSSR4}. It was even shown to arise in
opinion formation models \cite{SSSR5}.

The SSSR phenomenon occurs in extended systems, hence it is clearly
of great interest to describe this phenomenon within the NEP
framework. More, the NEP offers a general framework for the study of
the dependence of resonant and other related phenomena on any of
system's parameters. With such a goal in mind, in a recent paper
\cite{SSSR6} it was shown that SSSR could be analyzed within a NEP
framework and that, depending on the system, its origin could be
essentially traced back to a breaking of the symmetry of such a
potential. Even those cases discussed in \cite{SSSR4} could be
described within this same framework, and the (``effective") scaling
of the noise with the system's size could be clearly seen. Here, we
discuss in more detail the cases analyzed in \cite{SSSR6} and
present a new interesting one, corresponding to the study of SSSR in
a system that also shows noise induced patterns, the same one
studied in \cite{WeNew}. We show that in two of the cases
--corresponding to pattern forming systems that include only local
interactions-- the problem could be rewritten in such a way as to
present a kind of ``entrainment" between the symmetry breaking of an
``effective" potential together with a scaling of the noise
intensity with the system size.

The organization of the paper is as follows. In Section \ref{model1}
we focus on a simple reaction-diffusion model with a known form of
the NEP, that presents SSSR associated to a NEP's symmetry breaking.
In Section \ref{model2} we analyze the model of globally coupled
nonlinear oscillators discussed in \cite{SSSR4}, and show that it
can also be described within the NEP framework, with SSSR arising
due to a deepening of the potential wells, or through an
``effective" scaling of the noise intensity with the system size. We
start Section \ref{model3} by briefly reviewing the model and the
formalism to be used for the case of multiplicative noise. In this
case, by scaling the NEP with the system size, we show that the
system's behavior could be associated to a kind of ``entrainment"
between the symmetry breaking of an ``effective" potential and a
scaling of the noise intensity with the system size. Finally, we
present in Section \ref{conc} some conclusions and perspectives.

\section{\label{model1} A Simple Reaction-Diffusion System}

\subsection{Brief Review of the Model}

The specific model we shall focus on in this section, with a known
form of the NEP, corresponds to a one--dimensional, one--component
model \cite{WW,RL} that, with a piecewise linear form for the
reaction term, mimics general bistable reaction--diffusion models
\cite{WW}, that is with a cubic like nonlinear reaction term. In
particular we will exploit some of the results on the influence of
general boundary conditions (called {\it albedo}) found in
\cite{SW} as well as previous studies of the NEP \cite{I0} and of
SR \cite{extend2,extend2b,extend3a,extend3b}.

The particular non-dimensional form of the model that we work with
is \cite{SW,extend2,extend2b}
\begin{equation}
\label{Ballast} \frac{\partial}{\partial t }\phi = D\,\frac{\partial
^2}{\partial y^2} \phi - \phi + \phi_h \, \theta (\phi-\phi_c).
\end{equation}
We consider here a class of stationary structures $\phi(y)$ in the
bounded domain $y \in [-L,L]$ with albedo boundary conditions at
both ends, $$\left. {d\phi \over dy} \right|_{y=\pm L} = \mp k
\,\phi (\pm L),$$ where $k>0$ is the albedo parameter. It is worth
noting that for $k \rightarrow 0$ we recover the usual case of
Neumann boundary conditions (i.e. $\left. {d\phi \over dy}
\right|_{y=\pm L} = 0$), while for $k \rightarrow \infty$ what
results is the usual Dirichlet boundary conditions ($\phi (\pm L) =
0$).

Those stationary structures are the spatially symmetric (stable)
solutions to Eq.(\ref{Ballast}) already studied in \cite{SW}. The
explicit form of these stationary patterns is (see \cite{SW} for
details)
\begin{equation}\label{pattsw}
\phi (y)= \phi_{h} \, \left\{ %
\begin{array}{c}
\sinh (y_c/\sqrt{D}) \,\rho'(k,(L + y)/\sqrt{D}) \,
\rho(k,L/\sqrt{D})^{-1}, \,\,\,\,\,\,
\qquad \hbox{if } -L \leq y \leq -y_c, \\
1 - \cosh (y/\sqrt{D})\, \rho(k,(L - y_c)/\sqrt{D})\,
\rho(k,L/\sqrt{D})^{-1}, \qquad
\hbox{if} -y_c \leq y \leq y_c, \\
\sinh (y_c/\sqrt{D}) \, \rho'(k,(L - y)/\sqrt{D}) \,
\rho(k,L/\sqrt{D})^{-1}, \,\, \qquad \hbox{if } y_c \leq y \leq L,
\end{array}  \right.
\end{equation}
with $\rho(k,\zeta) = \sinh(\zeta)+k\ \cosh (\zeta)$, and
$\rho'(k,\zeta)= \frac{\partial \,\rho(k,\zeta)}{\partial\,\zeta}$.
The double-valued coordinate $y_c$, at which $\phi = \phi_c$, is
given by \cite{SW}
\begin{equation}
\label{yc} y_c^\pm = \frac{L}{2} - \frac{L}{2} \ln \left[ {z
\rho(k,L/\sqrt{D}) \pm \sqrt{z^2 \rho(k,L/\sqrt{D})^2 + 1 - k^2}
\over 1+k} \right],
\end{equation}
with $z=1-2\phi_c/\phi_h$ ($-1< z< 1$). Typical forms of the
patterns are shown in Fig. 1.

\begin{figure}
\centering
\resizebox{.6\columnwidth}{!}{\includegraphics{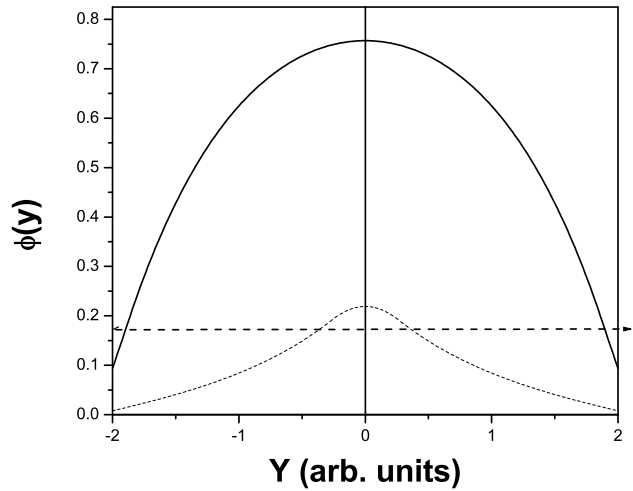}}
\label{fig0} \caption{Typical form of the patterns $\phi(y)$, for
$D=1.$, $L = 2$, $k = 2$ and $\phi_c/\phi_h = 0.193$. The continuous
line corresponds to the stable nonhomogeneous pattern, the dashed
one is the unstable pattern. In addition we also have the always
present, stable, null pattern.}
\end{figure}

When $y_c^\pm$ exists and $y_c^\pm < L$, this pair of solutions
represents a  structure with a  central ``excited'' zone
($\phi>\phi_c$) and two lateral ``resting'' regions ($\phi<\phi_c$).
For each parameter set, there are two  stationary solutions, given
by the two  values of $y_c$. Figure 5 in \cite{SW}, that we do not
reproduce here, depicts the curves corresponding to the relation
$y_c/L$ vs. $k$, for various values of $\phi_c/\phi_h$.

Through a linear stability analysis it has been shown \cite{SW} that
the structure with the smallest ``excited'' region (with
$y_c=y_c^+$, denoted by $\phi_u(y)$) is unstable, whereas the other
one (with $y_c=y_c^-$, denoted by $\phi_1(y)$) is linearly stable.
The trivial homogeneous solution $\phi_0(y)=0$ (denoted by $\phi_0$)
exists for any parameter set and is always linearly stable. These
two linearly stable solutions are the only stable stationary
structures under the given albedo boundary conditions. We will
concentrate on the region of values of $z$, $L$ and $k$, where both
nonhomogeneous structures exist.

For the system with the albedo b.c. that we are considering here,
the NEP reads \cite{I0}
\begin{equation}
\label{Lyap} {\cal F}[\phi,k,L] = \int_{-L}^{L} \left\{- \int_0
^{\phi(y,t)} \left[ -\phi'+\phi_h \theta(\phi'-\phi_c) \right] \
d\phi' +{D\over 2} \left( \frac{\partial}{\partial y}\phi(y,t)
\right) ^2 \right\} dy + \left.{k\over 2}\phi(y,t)^2\right|_{\pm L}.
\end{equation}
Strictly speaking, this is the system's Lyapunov functional, as we
are still considering the deterministic case. However, in what
follows we will always refer to the NEP both, for the deterministic
and stochastic cases. This functional fulfills the ``potential"
condition
\begin{equation}
\label{Ballast-d} \frac{\partial}{\partial t } \phi(y,t) = -
\frac{\delta}{\delta \, \phi(y,t)} {\cal F}[\phi,k,L],
\end{equation}
where $\frac{\delta}{\delta \,\phi(y,t)}$ indicates a functional
derivative.

Replacing the explicit forms of the stationary nonhomogeneous
solutions (Eq.(\ref{pattsw})), we obtain the explicit expression
\cite{extend2,I0}
\begin{equation}
\label{Lyapunov} {\cal F}^\pm = {\cal F}[\phi_{u,1},k,L] = -
\phi_h^2 \, y_c^{\pm} \, z +  \phi_h^2 \, \sinh(y_c^\pm/\sqrt{D}) \,
\frac {\rho(k,(L-y_c^\pm)/ \sqrt{D})}{\rho(k,L/\sqrt{D})},
\end{equation}
while for the homogeneous trivial solution $\phi_0=0$, we have
instead ${\cal F}[\phi_0,k,L]={\cal F}^0 = 0$.

Figure 2 depicts the nonequilibrium potential ${\cal F}[\phi,k,L]$
as a function of the system size $L$, for a fixed albedo parameter
$k$, and a fixed value of $\phi_c/\phi_h$ (that is, with fixed value
of $z$). The curves correspond to the NEP evaluated on the
nonhomogeneous structures, ${\cal F}^\pm$, whereas the horizontal
line stands for ${\cal F}^0$, the NEP of the trivial solution. We
have focused on the bistable zone, the upper branch being the NEP of
the unstable structure, where ${\cal F}$ attains a maximum, while in
the lower branch (for $\phi = \phi_0$ or $\phi = \phi_1$), the NEP
has local minima. We see that when $L$ becomes small, the difference
between the NEP for the states $\phi_u(y)$ and $\phi_1(y)$ reduces
until, for $L \approx 0.72$, they coalesce and, for even lower
values of $L$, disappear (inverse saddle-node bifurcation).

\begin{figure}
\centering
\resizebox{.6\columnwidth}{!}{\rotatebox{-90}{\includegraphics{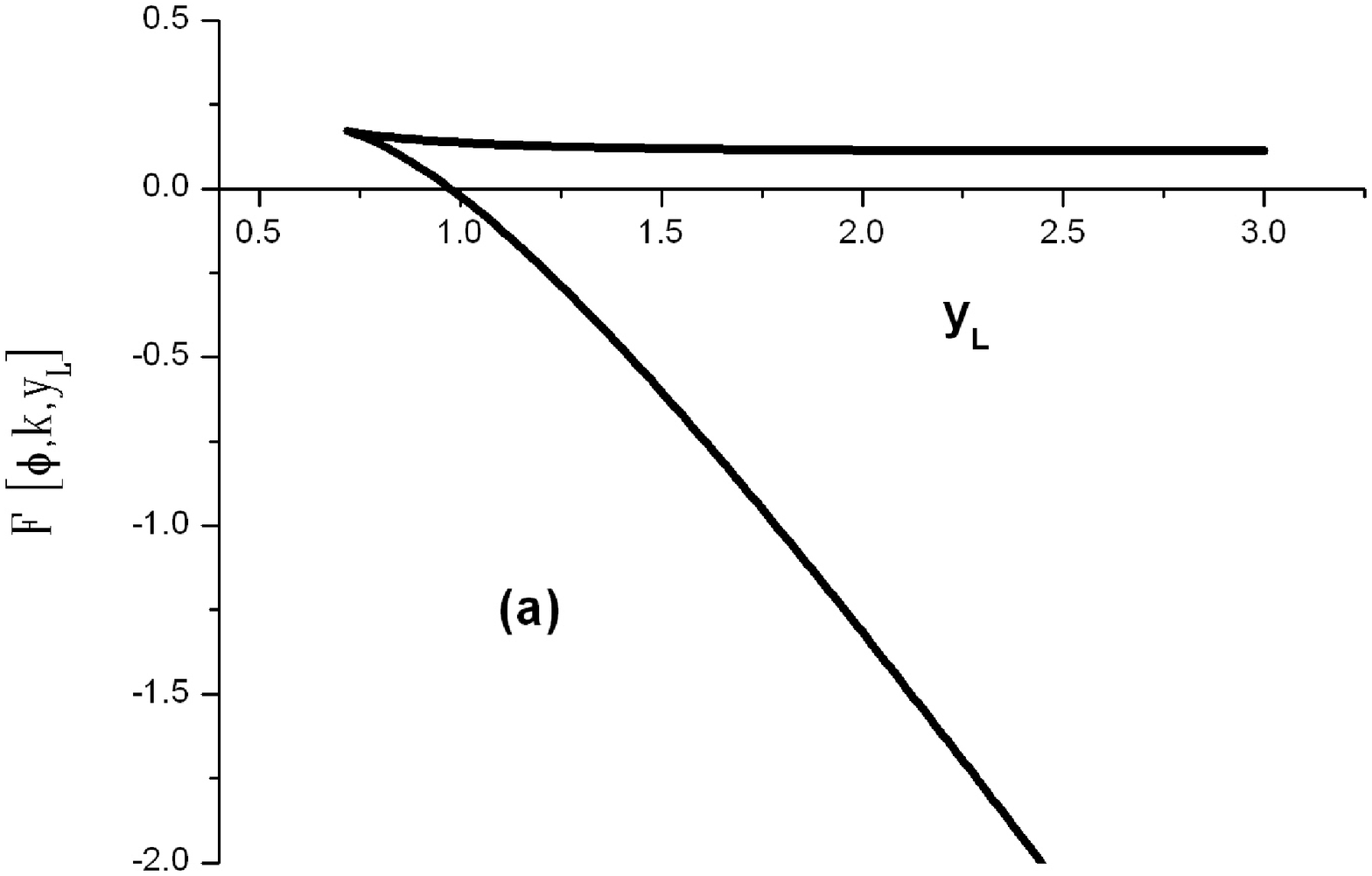}}}
\label{fig1a} \caption{${\cal F}[\phi,k,L]$ vs. $L$: NEP evaluated
at the stationary solutions $\phi_0(y)$, $\phi_1(y)$ and
$\phi_u(y)$. Here $k= 3$, $D=1.$, and $\phi_c/\phi_h = 0.193$.}
\end{figure}

It is important to note that, since the NEP for the unstable
solution $\phi_u$ is always positive and, for the stable
nonhomogeneous structure $\phi_1$, ${\cal F} < 0$ for $L$ large
enough, and ${\cal F} > 0$ for small values of $L$, the NEP for this
structure vanishes for an intermediate value $L = L^*$ of the system
size. At that point, the stable nonhomogeneous structure $\phi_1(y)$
and the trivial solution $\phi_0(y)$ exchange their relative
stability.

\begin{figure}
\centering
\resizebox{.6\columnwidth}{!}{\rotatebox{-90}{\includegraphics{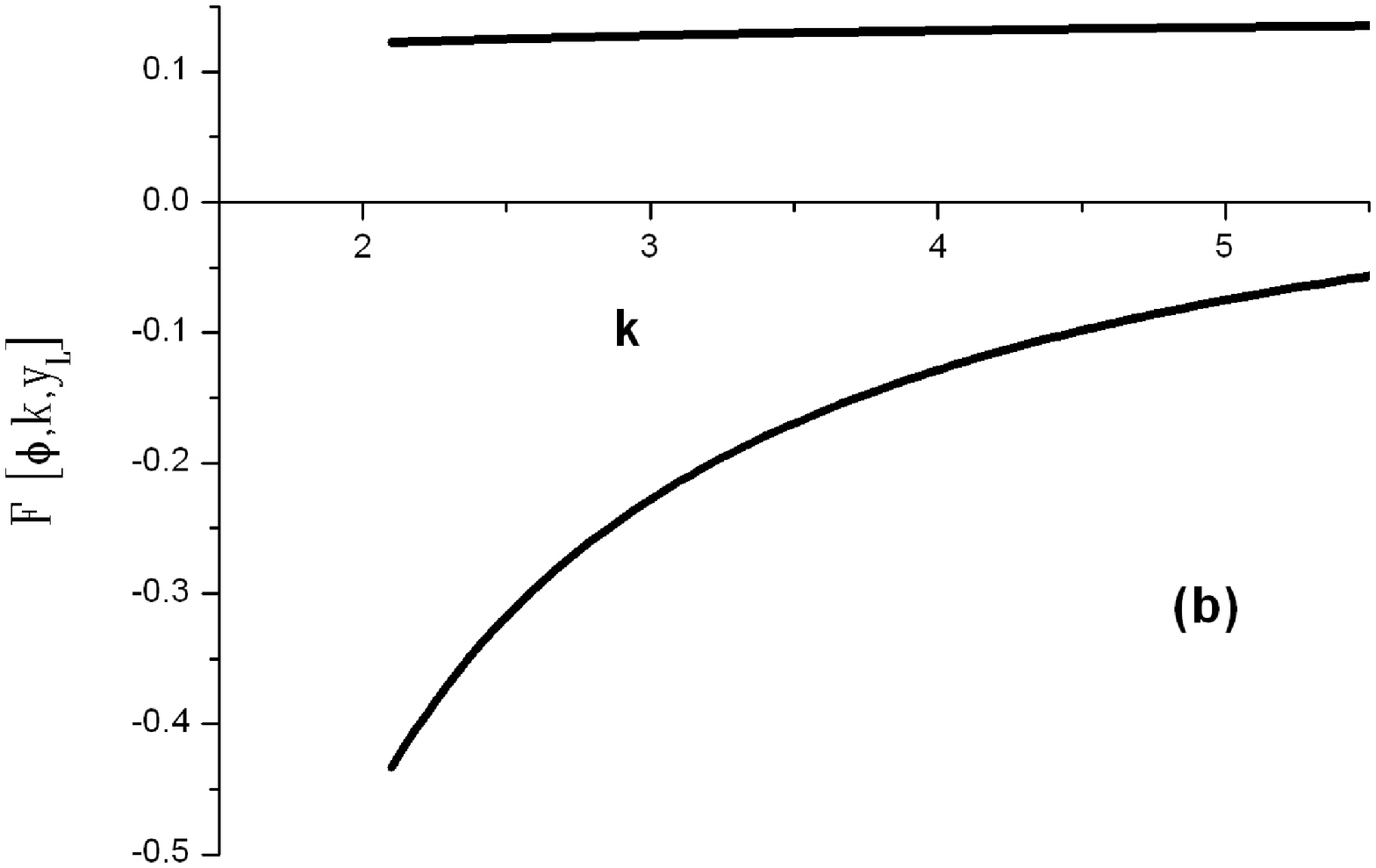}}}
\label{fig1b} \caption{${\cal F}[\phi,k,L]$ vs. $k$: NEP evaluated
at the stationary solutions $\phi_0(y)$, $\phi_1(y)$ and
$\phi_u(y)$. Here $L = 1.2$, $D=1.$, and $\phi_c/\phi_h = 0.193$.}
\end{figure}

For completeness and latter use, in Fig. 3 we show ${\cal
F}[\phi,k,L]$ but now as a function of $k$, for a fixed value of $L$
and the same value of $z$. Here we see that the initial large
difference between the NEP for the states $\phi_u(y)$ and
$\phi_1(y)$ reduces for increasing $k$ until, for $k \to \infty$,
the values for Dirichlet b.c. are asymptotically reached.

\subsection{System Size Stochastic Resonance}

In order to account for the effect of fluctuations, we include in
the time--evolution equation of our model (Eq.(\ref{Ballast})) a
fluctuation term, that we model as an additive noise source
\cite{extend3b,nsp}, yielding a stochastic partial differential
equation for the random field $\phi(y,t)$
\begin{equation}
\label{noise} \frac{\partial}{\partial t }\phi(y,t) = D\,
\frac{\partial ^2}{\partial y^2} \phi - \phi + \phi_h \, \theta
(\phi-\phi_c) + \xi(y,t).
\end{equation}
We make the simplest assumptions about the fluctuation term
$\xi(y,t)$, i.e. that it is a Gaussian white noise with zero mean
and a correlation function given by $$\langle\xi(y,t) \,
\xi(y',t')\rangle = 2~\gamma~\delta(t-t')~\delta(y-y'),$$ where
$\gamma$ denotes the noise strength.

As was discussed in \cite{extend2,extend2b,extend3a,extend3b}, using
known results for activation processes in multidimensional systems
\cite{HG}, we can estimate the activation rate according to the
following Kramers' like result for $\langle\tau\rangle$, the
first-passage-time for the transitions between attractors,
\begin{equation}
\label{tau} \langle\tau _{i}\rangle= \tau_{0} \, \exp \left\{ \frac{
\Delta {\cal F}^{i}[\phi,k]}{ \gamma } \right\},
\end{equation}
where $\Delta {\cal F}^{i}[\phi,k,L]={\cal F}[\phi_{u}(y), k, L] -
{\cal F}[\phi_{i}(y),k,L]$ ($i=0,1$). The pre-factor $\tau_{0}$ is
usually determined by the curvature of ${\cal F}[\phi,k,L]$ at its
extreme and typically is, in one hand, several orders of magnitude
smaller than the average time $\langle\tau\rangle$, while on the
other --around the bistable point-- does not change significatively
when varying the system's parameters. Hence, in order to simplify
the analysis, we assume here that $\tau_{0}$ is constant and scale
it out of our results. The behavior of $\langle\tau\rangle$ as a
function of the different parameters ($k$, $\phi_{c}/\phi_h$) was
shown in \cite{extend2,extend2b,I0}.

As was done in \cite{extend2}, we assume now that the system is
(adiabatically) subject to an external harmonic variation of the
parameter $\phi_{c}$: $\phi_{c}(t) = \phi_{c}^* + \delta \phi_{c}
\cos(\omega t)$ \cite{extend2b,extend3b}, and exploit the
``two-state approximation" \cite{RMP} as in
\cite{extend2b,extend3a,extend3b}. Such an approximation basically
consist in reducing the whole dynamics on the bistable potential
landscape to a one where the transitions occurs between two
states: the ones associated to the bottom of each well, while the
dynamics is contained only in the transition rates. For all
details on the general two-state approximation we refer to
\cite{extend3a}.

Up to first-order in the amplitude $\delta \phi_{c}$ (assumed to
be small in order to have a sub-threshold periodic input) the
transition rates $W_i$ adopt the form
\begin{equation} W_{i} =
\tau_0^{-1} \exp \left\{ - \frac{ \Delta {\cal F}^{i} [\phi,k,L,t]}{
\gamma } \right\}
\end{equation}
where
\begin{equation}
\label{modul} \Delta {\cal F}^{i}[\phi,k,L,t]=\Delta {\cal
F}^{i}[\phi,k,L] + \delta \phi_{c} \Bigl( \frac{\partial \Delta
{\cal F}^{i}[\phi,k,L]}{\partial \phi_{c}}
\Bigr)_{\phi_{c}=\phi_{c}^*} \cos (\omega t).
\end{equation}
This yields for the transition probabilities
\begin{equation}
\label{www} W_{i} \simeq \frac{1}{2} \Bigl(\mu_{i} \mp \alpha_{i}
\frac{\delta \phi_{c}}{\gamma} \cos(\omega t) \Bigr),
\end{equation}
where $$\mu_{i} \approx \exp\left\{ -\frac{\Delta{\cal F}^{i}
[\phi,k,L]}{\gamma}\right\} $$ and $$\alpha_{i} \approx \pm \mu_{i}
\left( \frac{d\Delta{\cal F}^{i}}{d\phi_{c}} \right) _{\phi_{c}^*},
$$
($i=1,2$). Using Eq. (\ref{Lyapunov}), it is clear that
$\frac{d\Delta{\cal F}^{i}}{d\phi_{c}}|_{\phi_{c}^*}$ can be
obtained analytically.

These results allows us to calculate the autocorrelation function,
the power spectrum density and finally the SNR, that we indicate by
$R$. The details of the calculation were shown in \cite{extend3a}
and will not be repeated here. For $R$, and up to the relevant
(second) order in the signal amplitude $\delta \phi_{c}$, we obtain
\cite{extend3a}
\begin{equation}
\label{snr} R= \, \frac{\pi}{4\,  \mu_1 \, \mu_2} \frac{(\alpha_2 \,
\mu_1 + \alpha_1 \, \mu_2)^2}{\mu_1 + \mu_2}.
\end{equation}
Due to the form of $\alpha_{i}$, we can reduce the previous
expression to
\begin{equation}
\label{snrp} R= \, \frac{\pi}{4\, \gamma ^2} \frac{\mu_1 \,
\mu_2}{\mu_1 + \mu_2} \Phi,
\end{equation}
where
\begin{equation} \Phi = \left[ \int_{-L/2}^{L/2} \phi_h \,
\theta (\phi_{st}(y)-\phi_c) \, dy \right]^2 = \left[ 2 \, \phi_h \,
y_c(L) \right]^2.
\end{equation}
We have now all the elements required to analyze the problem of
SSSR.

Figure 4 shows the typical behavior of SR, but now --in the
horizontal axis-- the noise intensity is replaced by the the system
length $L$, for fixed values of $k$, $\gamma$ (the noise intensity)
and the ratio $\phi_c/\phi_h$ (that in our scaled system is a single
parameter). Such a response is the expected one for a system
exhibiting SSSR. Within the context of NEP, it results clear that,
in this kind of systems, the phenomenon arises due to the breaking
of the NEP's potential symmetry. This means that, as shown in Fig.
2, due to the variation of $L$, both attractors can exchange their
relative stability. For a value $L = L^*$, both stable structures,
the nonhomogeneous $\phi_1(y)$ and the trivial $\phi_0(y)$, have the
same value for the NEP. When $L < L^*$, $\phi_1(y)$ becomes less
stable than $\phi_0(y)$, making the transitions from $\phi_1(y)$ to
$\phi_0(y)$ ``easier" (the barrier is lower) than in the reverse
direction, reducing the system's response. When $L \sim 0.72$,
$\phi_1(y)$ and $\phi_u(y)$ coalesce and disappear, and the response
is strictly zero (within the linear response implicit in the two
state approximation). When $L > L^*$, $\phi_1(y)$ becomes more
stable than $\phi_0(y)$, making now the transitions from $\phi_0(y)$
to $\phi_1(y)$ ``easier" than in the reverse direction, again
reducing the system's response. Clearly, the system's response has a
maximum when both attractors have the same stability ($L = L^*$),
and decays when departing from that situation. Hence, for this
system and within this framework, SSSR arises as a particular case
of the more general discussion done in \cite{extend3a}.

\begin{figure}
\centering
\includegraphics[width=7cm,angle=-90]{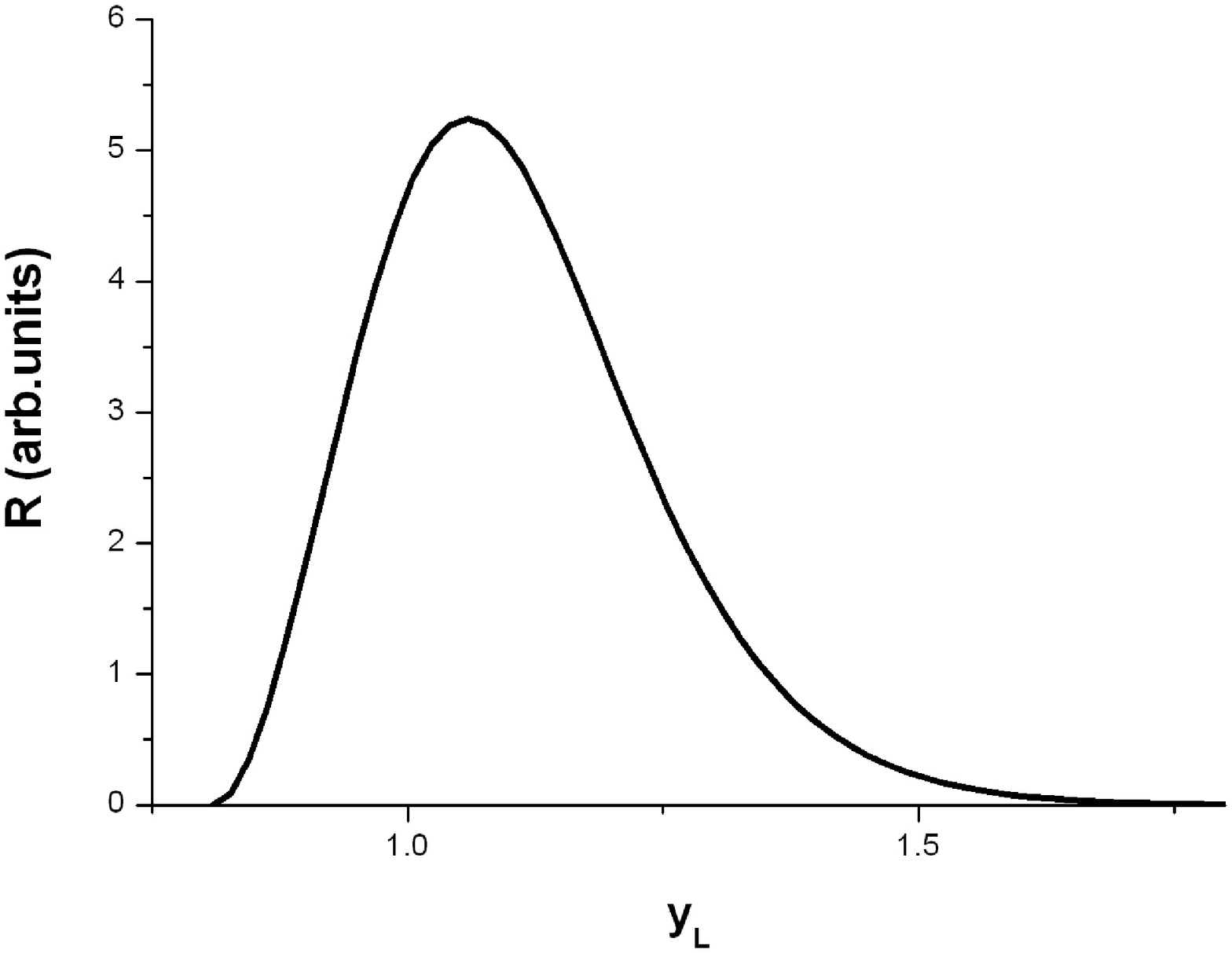}
\label{fig2} \caption{SNR, vs. $L$, for $k = 3.$, $\gamma = 0.1$,
$D=1.$, and $\phi_c/\phi_h = 0.193$.}
\end{figure}

We can analyze the same problem from an alternative point of view.
That is, studying the scaling of the NEP in Eq. (\ref{Lyap}) with
$L$. Due to the similarities of the present problem with the one
discussed in Section \ref{model3}, we stop here the discussion, and
left such a kind of analysis to treat that problem.

To conclude this section as well as for completeness, we change the
point of view. In Fig. 5 we show the curves of the SNR as a function
of $k$, while keeping fixed values of $L$, and $z$. When $k$ is not
too large, indicating a high degree of reflectiveness at the
boundary (that is, a reduced exchange with the environment), we see
that the SNR changes for $k$ varying from low to larger values,
showing a broad resonance like curve. Remember that a large value of
$k$ indicates that the system boundaries become absorbent. Such a
broadening of the resonance indicates the {\em robustness} of the
systems' response when varying $k$, a parameter that somehow
indicates a degree of coupling with the environment. However,
according to the previous argument --about the breaking of NEP's
symmetry-- from the behavior in Fig. 3 this is again the expected
result.

\begin{figure}
\centering
\resizebox{.6\columnwidth}{!}{\includegraphics[angle=90]{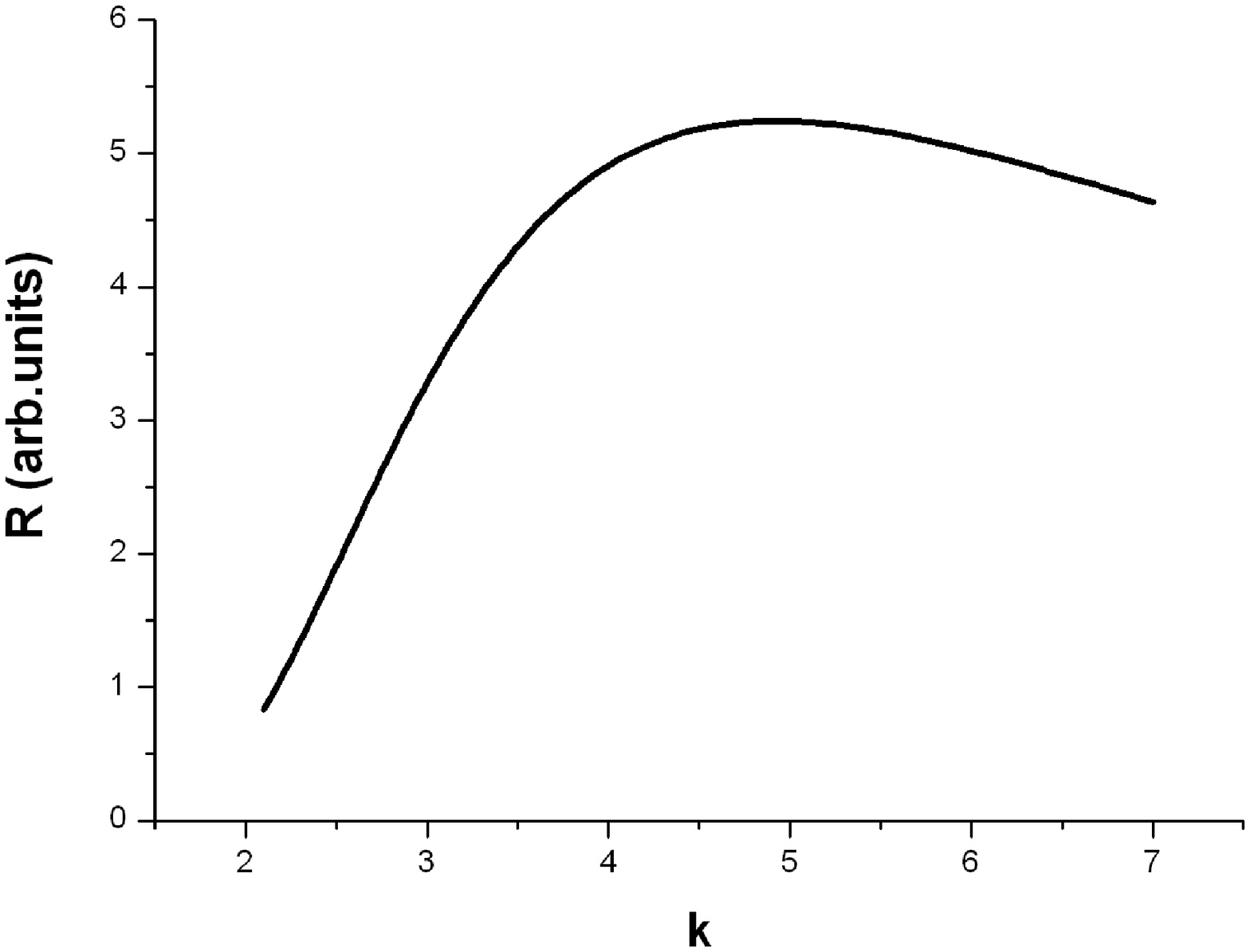}}
\label{fig3} \caption{SNR, vs. $k$ for $L = 1.2$, $\gamma = 0.1$,
$D=1.$, and $\phi_c/\phi_h = 0.193$.}
\end{figure}

\section{\label{model2} The Global Coupling Model}

In this section we consider one of the models discussed in
\cite{SSSR4} from the point of view of the NEP approach. The model
we refer to is described by a set of (globally) coupled nonlinear
bistable oscillators
\begin{eqnarray}
\label{PIKOV1} \dot{x}_{j} & = & x_{j} - x_{j}^3 +
\frac{\varepsilon}{N} \sum_{k=1}^{N} (x_{k}-x_{j}) +
\sqrt{2\,\gamma}\, \xi_{j}(t) + f_j(t),\nonumber \\
\dot{x}_{j} & = & - \frac{\partial}{\partial \,x_{j}} U(\{ x \},t) +
\sqrt{2\,\gamma}\, \xi_{j}(t),
\end{eqnarray}
with $f_j(t) = A \cos (\omega t)$, $\{ x \} = (x_{1},x_{2},..
,x_{N})$, $\xi_{j}(t)$ are Gaussian noises with zero mean and
$\langle \xi_{j}(t) \xi_{l}(t') \rangle = \delta_{jl} \,
\delta(t-t')$, and where we have defined
\begin{eqnarray}
\label{PIKOV2} U(\{ x \},t) & = & U_0(\{ x \}) - A \cos (\omega t)
\sum_{j=1}^{N} x_{j} \nonumber \\
& = & \sum_{j=1}^{N} \left( \frac{x_{j}^4}{4} - \frac{x_{j}^2}{2}
\right) + \frac{\varepsilon}{2 \, N} \sum_{j=1}^{N} \sum_{k=1}^{N}
(x_{k}-x_{j})^{2} - A \cos (\omega t) \sum_{j=1}^{N} x_{j}
\nonumber \\
& = & \sum_{j=1}^{N} u_0(x_j) + \frac{\varepsilon}{2 \, N}
\sum_{j=1}^{N} \sum_{k=1}^{N} (x_{k}-x_{j})^{2} - A \cos (\omega t)
\sum_{j=1}^{N} x_{j}.
\end{eqnarray}
Due to the structure of Eq. (\ref{PIKOV1}) it is clear that
$U_0(\{ x \})$, the potential function in Eqs.
(\ref{PIKOV1},\ref{PIKOV2}), is the discrete form of the NEP for
this problem. For $A=0$ the stationary distribution of the
multidimensional Fokker-Planck equation associated to Eq.
(\ref{PIKOV1}) results
\begin{equation}
\label{PIKOV3} P_{stat}(\{ x \}) \approx \exp \left(- \frac{U_0(\{
x \})}{\gamma} \right).
\end{equation}
This potential has two attractors corresponding to $x_{1}=x_{2}=...
=x_{N}=\pm 1$, and a barrier separating them at $x_{1}=x_{2}=...
=x_{N}=0$.

Now, exploiting the same scheme as before but, as both attractors
have the same ``energy", reduced to the symmetric case, we get for
the SNR
\begin{equation}
\label{PIKOV4} R \approx \exp \left(- \frac{\triangle U_0(\{ x
\})}{\gamma} \right) \approx \frac{N}{\gamma} \, \exp \left(-
\frac{N\, \triangle u_0(X)}{\gamma}\right),
\end{equation}
where $X$ is a kind of ``collective coordinate" (the one evolving
along the trajectory joining both attractors, that pass though the
saddle, and that can be approximately interpreted as $X \approx
\frac{1}{N}\sum_{j=1}^{N} x_{j}$. However, we need the evaluation at
only two points: $X=0, \pm 1$), and
$$\triangle u_0(X) = u_0(X = \pm 1) - u_0(X = 0).$$
Such a SNR clearly shows similar SSSR characteristics as those
described in \cite{SSSR4}. In this situation the NEP's symmetry is
retained when varying $N$, while the wells are deepened (or the
barrier separating them is enhanced). However, if we scaled out $N$,
we find a constant ``effective" potential ($u_0(X)$), while the
system shows an {\it effective} scaling of the noise with $N$. In
this case we could speak of a {\em noise scaled} SSSR, in contrast
to the previous case that could be called a {\em NEP symmetry
breaking} SSSR.

In order to deepen our understanding of this case, let us analyze
a continuous model, that is tightly connected with the previous
discrete one. Consider a field $\psi (y,t)$, that behaves
according to the following functional equation
\begin{eqnarray}
\label{PIKOV1p} \frac{\partial}{\partial \,t}\psi (y,t) & = &
\psi(y,t) - \psi(y,t)^3 + \varepsilon \int_{\Omega}
(\psi(y',t)-\psi(y,t))\,dy' + \xi(y,t) + f(t),\nonumber \\
& & \nonumber \\
& = & - \frac{\delta}{\delta \,\psi(y,t)} U(\psi(y,t)) + \xi(y,t),
\end{eqnarray}
with $f(t) = A \cos (\omega t)$, while for the noise we assume
that, as before, it is white and Gaussian with $\langle \xi(y,t)
\rangle = 0$, and the correlation
$$\langle \xi(y,t) \xi(y',t') \rangle = 2\, \gamma \, \delta(y-y')
\delta(t-t').$$ $\Omega$ indicates the integration range, and
$\frac{\delta}{\delta \,\psi(y,t)}$ is a functional derivative. We
consider a finite system in the interval $y \in [-L/2,L/2]$, and
assume Neumann boundary conditions. The form of the potential
$U(\psi(y,t),t)$ results
\begin{eqnarray}
\label{PIKOV2p} U(\psi(y,t)) & = & U_0(\psi(y,t)) - F(\psi(y,t))
\nonumber \\
&  & \nonumber \\
& = & \int_{\Omega}\, dy \, u_0(\psi(y,t)) + \frac{\varepsilon}{2}
\,\int_{\Omega}\,dy\,\int_{\Omega}\,dy'\,
\left( \psi(y',t)-\psi(y,t) \right) ^{2} - F(\psi(y,t)),\nonumber \\
&  &
\end{eqnarray}
where $$u_0(\psi(y,t)) = \left( \frac{\psi(y,t)^4}{4} -
\frac{\psi(y,t)^2}{2} \right),$$ and $$F(\psi(y,t)) = A \cos
(\omega t) \int_{\Omega} \psi(y,t) \, dy.$$ This potential is
clearly similar to the one discussed in \cite{extend2b}, but with
the local (diffusive) coupling being zero, and the nonlocal
contribution becoming ``global". For $A=0$ the stationary
distribution of the multidimensional Fokker-Planck equation
associated to Eq. (\ref{PIKOV1p}) results
\begin{equation}
\label{PIKOV3f} P_{stat}(\psi_{stat}(y)) \approx \exp \left(-
\frac{U_0(\psi_{stat}(y))}{\gamma} \right),
\end{equation}
with $\gamma$ the noise intensity. This potential has two
attractors corresponding to the constant fields $\psi_{stat}(y)=
\pm 1$, and a barrier separating them at $\psi_{stat}(y)=0$.

Hence, exploiting the same scheme as in the previous section, but
reduced to the symmetric case as both attractors have the same
``energy", we get for the SNR
\begin{equation}
\label{PIKOV4p} R \approx \exp \left(- \frac{\triangle
U_0(\psi_{stat}(y))}{\gamma} \right) \approx \frac{L}{\gamma} \,
\exp \left(- \frac{L\, \triangle u_0(\psi_{stat}(y))} {\gamma}
\right).
\end{equation}
where $$\triangle u_0(\psi_{stat}(y)) = u_0(\psi_{stat}(y) = \pm
1) - u_0(\psi_{stat}(y)= 0).$$ This SNR clearly shows the same
SSSR characteristics as those described for the discrete case,
where the role of $N$ (number of elements) is now played by $L$
(size of the system).

Note that Eq. (\ref{PIKOV1p}) corresponds to the continuous limit of
Eq. (\ref{PIKOV1}) and  the result for the discrete case (Eq.
(\ref{PIKOV4})) is recovered in Eq. (\ref{PIKOV4p})  for the
normalized noise intensity $\gamma_{dc} =\gamma / \Delta x$ (with
$\gamma_{dc}$ the noise intensity for the discrete case), and $L=N
\Delta x.$

To conclude this section, we refer to another case discussed in
\cite{SSSR4}, the one corresponding to the Ising model. Such a case
has many similarities with the case of the set of coupled nonlinear
bistable oscillators discussed above. It can be described in a
similar way to the case above. That means we could also find an
effective potential playing the role of the NEP, having two
attractors (corresponding to all spins up or all down), a barrier
corresponding to a mixed state, whose high depends linearly with $N$
(the number of spins). The final result will be similar to the one
in Eq. (\ref{PIKOV4}) above.

\section{\label{model3} Multiplicative Noise Case}

\subsection{Brief Review of the Model}

The basic model to be considered in this section is the same one
studied in \cite{WeNew}, and consist of the following ensemble of
nonlinear coupled  oscillators, described in terms of a continuous
field
\begin{equation} \label{master}
\frac{\partial}{\partial t} \phi(y,t) = \frac{\partial}{\partial
y} \left( D(\phi) \, \frac{\partial}{\partial y} \phi \right) +
f(\phi)+ \frac{1} {\sqrt{D(\phi)}} \, \xi(y,t).
\end{equation}
Here $\xi(y,t)$ is again a Gaussian white noise with zero mean and
correlation $\langle \xi(y,t) \xi(y',t') \rangle=2 \gamma
\delta(y-y') \delta(t-t')$, being $\gamma$ the noise intensity.
$D(\phi)$ is a field-dependent diffusivity and the coefficient of
the noise term guarantee that fluctuation-dissipation  relation is
fulfilled \cite{kitara}. The nonlinearity $f(\phi)$ which drives
the dynamics in absence of noise is monostable, and we adopt a
density dependent diffusion coefficient to generate a
noise-induced bistable dynamics. In particular, we use
\begin{equation} \label{difu}
D(\phi)=\frac{D_0}{1+h \phi^2 },
\end{equation}
and
\begin{equation}
f(\phi)=-\phi^3+b \, \phi,
\end{equation}
being $D_0$, $h$ and $b$  positive constants. We will consider a
finite system, limited to ($-L/2 \le x \le L/2$), and assume
Dirichlet boundary conditions $\phi(\pm L/2)=0$.

As we are considering the Stratonovich interpretation, the
stationary solution  of the probability $P_{st}(\phi)$ of the
stochastic field $\phi(x,t)$ given by  Eq. (\ref{master}) can be
written \cite{ibanyes} in terms of an effective potential
\begin{equation} \label{solucion}
P_{st}(\phi) \sim \exp(-V_{eff}/\gamma),
\end{equation}
with
\begin{eqnarray} \label{poteff}
V_{eff}[\phi] & = & \int_{-L/2}^{L/2} dy \bigg \{\frac{1}{2} \left(
D(\phi)\frac{\partial }{\partial y} \phi \right) ^2-U(\phi) -\lambda
\, \ln D(\phi) \bigg \}, \\
& & \nonumber
\end{eqnarray}
(where \(U(\phi)=\int_0^{\phi} D(\phi') f(\phi')\,d\phi'\)). Here
$\lambda$ is a renormalized parameter, related to $\gamma$ by
$\lambda=\gamma/2\,\Delta y$ in a square discrete lattice, where
$\Delta y$ is the lattice parameter \cite{ibanyes}. The extremes
of $V_{eff}$--- stationary noise-induced structures of the
effective dynamics--- can be computed from
\begin{equation}
\label{equivalente} \frac{\partial }{\partial y} \left( D(\phi) \,
\frac{\partial }{\partial y} \phi \right) +F_{eff}(\phi) =0
\end{equation}
with an effective nonlinearity
\begin{equation}
\label{fundamental} F_{eff}(\phi)=f(\phi)+\lambda
\frac{1}{D(\phi)^2}\frac{d}{d\phi} D(\phi)=\phi
\,(\phi-\phi_1)\,(\phi_2-\phi),
\end{equation}
where $\phi_{1,2}$ depend on parameters, in particular on the
renormalized noise intensity $\lambda$. (We have found one trivial
homogeneous structure $\phi=0$  and two nonhomogeneous patterns,
the unstable (saddle) $\phi_u$ and the stable one $\phi_s$ (see
Ref. \cite{WeNew})).

We note that in the deterministic problem we have a monostable
\cite{foot1} reaction term ($\lambda=0$). As we increase the noise
intensity, due to the noise effects, we have an effective nonlinear
term $F_{eff}$ that results to be bistable (in the interval
$0<\lambda<D_0/(2h)$) and finally monostable for $\lambda>D_0/(2h)$
(reentrance effect). We also remark that $\phi=0$ is always a root
of $F_{eff}$ (see Fig. 6), and it is an extremum of $V_{eff}[\phi]$
for all values of \(\lambda\). In what follows we will call this
structure $\phi_0$. As a final remark, the situation here is that
the same noise that induces the patterns and the bistability is the
one inducing the transitions among them and the SR phenomenon.

\begin{figure}
\centering
\includegraphics[width=10cm]{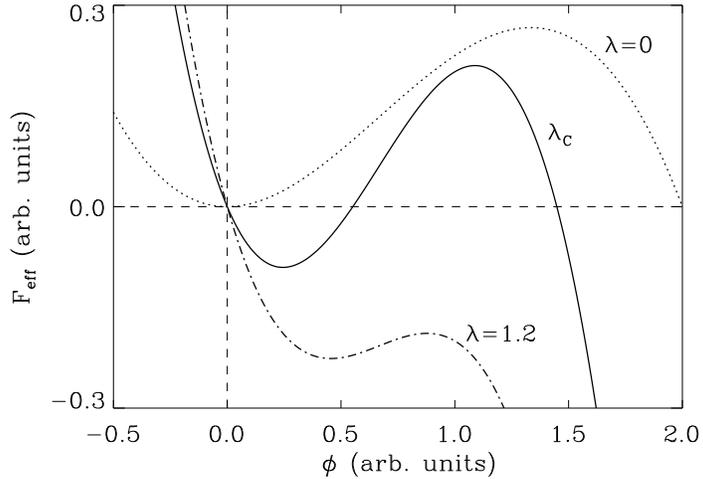}
\label{fignonlin}\caption{Form of the nonlinearities for the
deterministic case ($\lambda=0$), bistable case ($\lambda=0.8$) and
a monostable case ($\lambda=1.2$) in the reentrance region. The
vertical scale was changed in the deterministic case in order to
clarify the figure. Note that $\phi=0$ remains as a root in all
cases. The parameters used are: $D_0=1$, $h=1/2$ and $b=2$.}
\end{figure}

\subsection{\label{sr:size} System Size Stochastic Resonance}

To analytically describe the stochastic resonance, again we resort
to use the two state approach in the adiabatic limit \cite{RMP}.
As indicated before, all details about the procedure and the
evaluation of the SNR could be found in \cite{extend3a}. The
system is now subject to a time periodic subthreshold signal
$b=b_0+ S(t)$ where $S(t)=\Delta b \, \sin(\omega_0 t )$. Up to
first-order in the small amplitude \(\Delta b\) the transition
rates $W_i$ take the form
\begin{eqnarray}
\label{rates1} W_1(t) & = & \mu_1 - \alpha_1 \,\Delta b\,
\sin(\omega_0 \, t), \nonumber\\
                W_2(t) & = & \mu_2 + \alpha_2 \,\Delta b\,
\sin(\omega_0 \, t),
\end{eqnarray}
where the constants $\mu_{1,2}$ and $\alpha_{1,2}$ are obtained from
the Kramers-like formula for the transition rates \cite{HG}
\begin{equation} \label{rates}
W_{\phi_i \rightarrow \phi_j}= \frac{\beta_+}{2 \pi} \,
\left[\frac{\det \,V_{eff}(\phi_i)}{\vert \det \,V_{eff}(\phi_s)
\vert} \right]^{1/2} \, \exp[- (V_{eff}(\phi_i)-
V_{eff}(\phi_s))/\gamma].
\end{equation}
Here $\beta_+$ is the unstable eigenvalue of the deterministic
flux at the relevant saddle point ($\phi_s)$ and
\begin{eqnarray}
\mu_{1,2} & = & W_{1,2} \vert_{S(t)=0},     \nonumber \\
\alpha_{1,2} & = & \mp  \frac{dW_{1,2}}{dt}\vert_{S(t)=0}.
\end{eqnarray}

We note in passing that, due to the system's sensitivity to small
variations in the parameters, and at variance with the case
studied in Section II, here we require the evaluation of the
pre-factor in Eq. (\ref{rates}).

As before, these results allows us to calculate the autocorrelation
function, the power spectrum and finally the SNR (indicated by
\(R\)). For \(R\), up to the relevant (second) order in the signal
amplitude \(\Delta b\), similarly to Eq. (\ref{snr}) and
(\ref{snrp}) we obtain
\begin{equation} \label{sigtonoi}
R=\frac{\pi}{4\,  \mu_1 \, \mu_2} \frac{(\alpha_2 \, \mu_1+\alpha_1
\, \mu_2)^2}{\mu_1 + \mu_2}= \frac{\pi}{4 \gamma ^2} \, \frac{\mu_1
\, \mu_2}{\mu_1+\mu_2} \, \Phi,
\end{equation}
where now
\begin{equation} \label{spatial}
\Phi=\int_{-L/2}^{L/2} dy \, \int_{\phi_0}^{\phi_s(y)} \, D(\phi')
\,  \phi'^2 \, d\phi'
\end{equation}
gives a simultaneous measure of the spatial coupling (through
\(D(\phi)\)) and the system size extension (through $\int dy$). In
our previous work \cite{WeNew} we have found that the dependence
of the SNR as a function of \(\lambda\) is maximum at the
symmetric situation \(\lambda=\lambda_c = 0.8\), where both stable
structures (\(\phi_0\) and \(\phi_s\)) have the same stability
(\(V_{eff}[\phi_0]=V_{eff}[\phi_s]\)).

To analyze the system size dependence of the SR, we fix the
renormalized noise intensity (\(\lambda=\lambda_c = 0.8\)) and the
parameters \(D_0=1\), \(b=2\) and \(h=1/2\); only change the
length \(L\) (with fixed lattice parameter \(\Delta x\)).

It is worth here remarking that in \cite{SSSR4} what is varied is
the length of the lattice, while the noise intensity, the
coupling, etc, are kept constant. In the present case, due to the
characteristics of the model, we have that as $L$ is varied, the
fields change inducing the change of the diffusive coupling,
making a strong difference with the case in \cite{SSSR4}.

At this point we can just analyze the dependence of the SNR with $L$
as was done in Section \ref{model1}. However, as was indicated near
the end of that section, we use an alternative for of analysis,
looking at the scaling of the potential with $L$. We consider the
following transformations
\begin{itemize}
\item \(y \to x=\frac{y}{L}\), \item \(D_o \to
D_1=\frac{D_o}{L}\), \item \(D(\phi)=\frac{D_o}{1+h\,\phi^2} \to
D_1(\phi)=\frac{D_1}{1+h\,\phi^2}\),
\end{itemize}
the effective potential (Eq. (\ref{poteff}) could be written as
\begin{equation}
V_{eff}[\phi]=L\,\int_{-1/2}^{1/2} dx \bigg \{ -U(\phi)+ \frac{1}{2}
\left[ \frac{D(\phi)}{L} \frac{\partial}{\partial x} \phi \right] ^2
\bigg \} -\lambda\,L \int_{-1/2}^{1/2} dx \ln D(\phi),
\end{equation}
that finally yields (using the previous definition
\(D(\phi)/L=D_1(\phi)\))
\begin{equation}\label{vefectivo}
V_{eff}[\phi]/L=V_{sc}(L)=\,\int_{-1/2}^{1/2} dx \bigg \{ -U(\phi)+
\frac{1}{2} \left[ D_1(\phi)\frac{\partial}{\partial x} \phi \right]
^2 -\lambda \ln D_1(\phi) \bigg \} - \lambda \ln L.
\end{equation}
Here it becomes apparent that this scaling yields a logarithmic
length contribution to the scaled potential.

The stationary solution \(P_{st}(\phi)\) (see Eq. \ref{solucion}) of
the stochastic field $\phi(y,t)$ can be written
\begin{equation}
P_{st}(\phi) \sim \exp(-V_{sc}(L)/\gamma_x(L)),
\end{equation}
with \(\gamma_x(L)=\gamma/L\).

We can also consider the scaling of the spatial factor \(\Phi\) (see
Eq. (\ref{spatial})). It results
\begin{equation}
\Phi = L \, \int_{-1/2}^{1/2} dx \, \int_{\phi_0}^{\phi_s(x)}
D(\phi') \,  \phi'^2 \, d\phi = L \, \frac{D_0}{h}
\int_{-1/2}^{1/2} dx \,
\left\{\phi_s(x)-\frac{\arctan(\sqrt{h}\,\phi_s(x))}{\sqrt{h}}\right\}.
\end{equation}
Hence, we have the dependence of the NEP as well as transitions
rates (Eq. (\ref{rates})) and finally of the SNR (Eq.
(\ref{sigtonoi})), on the system length (or the number of coupled
elements for discrete systems). To illustrate this, in Fig. 7 we
show $V(\phi_{eff})$ as a function of $L$. The behavior shown in
this figure is analogous to the one observed in Fig. 2. Hence, we
can anticipate the existence SSSR in this system.

We see that for small $L$, small size effects increase the NEP
values of the nonhomogeneous structures, and the uniform state
results to be the most stable one. It becomes metastable at $L_c$,
and nonuniform patterns are the globally stable attractors for
larger values of $L$. The  rate transitions also reflect this fact
(they are decreasing functions of $L$). However, we expect that
$\Phi$--- which depends on the system size---, increases with $L$,
and due to the interplay of the rates and the behavior of $\Phi$, a
SSSR can be expected. Such a behavior becomes apparent in Fig. 8.

\begin{figure}
\centering
\includegraphics[width=10cm]{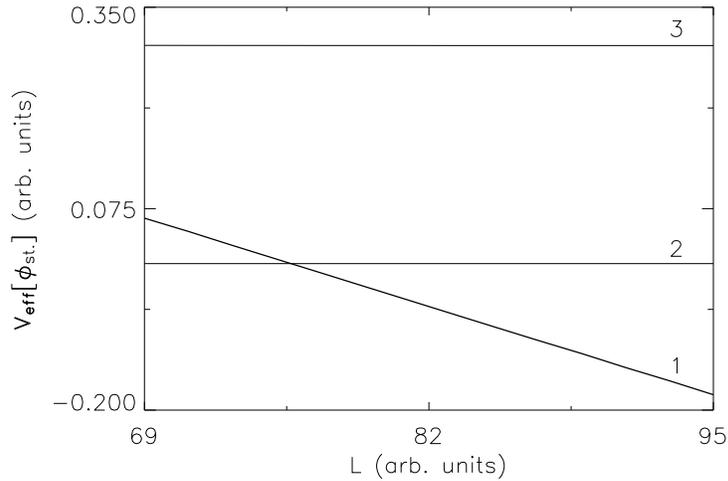}
\label{figpot} \caption{Nonequilibrium potential
\(V_{eff}[\phi_{st}]\) evaluated in the stationary structures as a
function of the system size \(L\). Curves correspond to: (1)
stable (\(\phi_s\)), (2) homogeneous (\(\phi_0\)) and (3) unstable
(\(\phi_u\)) patterns. Note the global stabilization of the
nontrivial stable pattern for high values of \(L\).}
\end{figure}

\begin{figure}
\centering
\includegraphics[width=10cm]{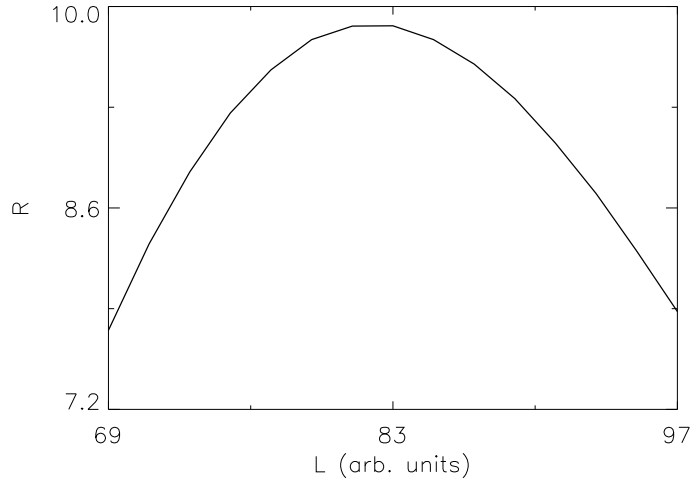}
\label{figSSSR}\caption{Signal-to-noise ratio vs. $L$. As
indicated in the text, we fixed $D_o=1$, $h=0.5$,  $b=2$, and
$\lambda=\lambda_c=0.8$.}
\end{figure}

We can make the following interpretation in terms of $V_{sc}(L)$,
the ``effective" NEP (the scaled form of the NEP, Eq.
(\ref{vefectivo})), and the the scaling of the noise intensity. In
Fig. 9 we depict the form of $V_{sc}(L)$ as function of $L$. It is
clear that, even though it is weak, the dependance of $V_{sc}(L)$
with $L$ still shows the change in the relative stability of the
attractors, while we have the scaling of the noise intensity with
$L$. Hence, we can argue that there is a kind of ``effective
entanglement" between the symmetry breaking and the noise scaling.

\begin{figure}
\centering
\resizebox{.6\columnwidth}{!}{\includegraphics{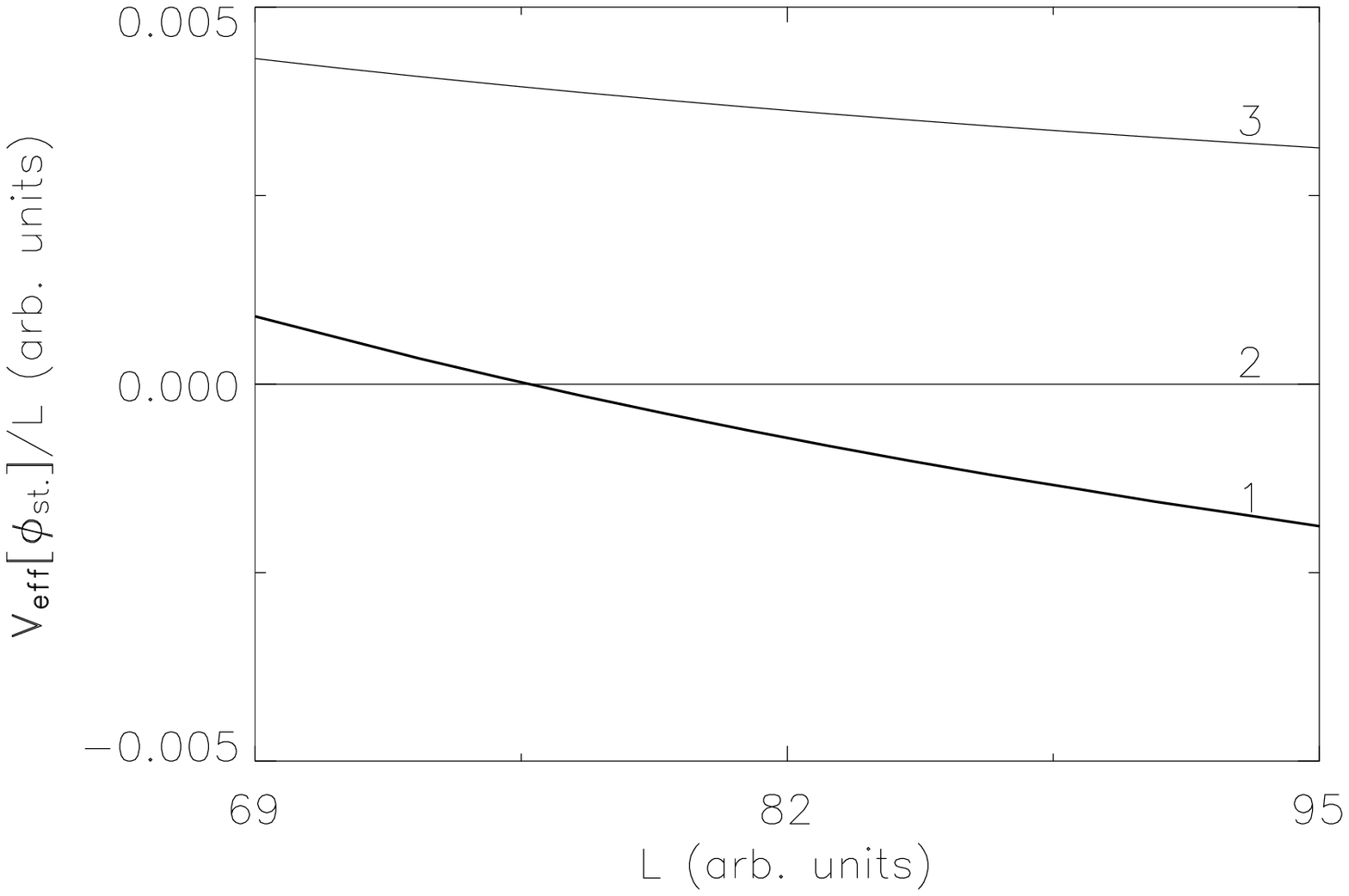}}
\label{figpotcs} \caption{$V_{sc}(L)/L$ evaluated on the stationary
structures as a function of the system size \(L\). Curves correspond
to: (1) stable (\(\phi_s\)), (2) homogeneous (\(\phi_0\)) and (3)
unstable (\(\phi_u\)) patterns. These corresponds to the same curves
as in Fig. 7. }
\end{figure}

\section{\label{conc}Conclusions}

The study of SR in extended or coupled systems, motivated by both,
some experimental results and the growing technological interest,
has recently attracted considerable attention
\cite{extend1,otros,extend2,extend2b,extend3a,extend3b,extend3c}.
In previous papers
\cite{extend2,extend2b,extend3a,extend3b,extend3c} we have studied
the SR phenomenon for the transition between two different
patterns, exploiting the concept of \textit{nonequilibrium
potential} \cite{GR,I0,WeNew}.

Recent works \cite{SSSR1,SSSR2,SSSR3} have shown that several
systems presents intrinsic SR-like phenomena as the number of units,
or the size of the system is varied. This phenomenon, called
\textit{system size stochastic resonance}, has also been found in a
set of globally coupled units described by a $\phi ^{4}$ theory
\cite{SSSR4}, and even shown to arise in opinion formation models
\cite{SSSR5}. Such SSSR phenomenon occurs in extended systems, hence
it was clearly of great interest to have a description of this
phenomenon within the NEP framework.

Here, we have discussed in detail two of the cases analyzed in
\cite{SSSR6} and presented a third, interesting one, that
corresponds to the study of the system size dependence of SR in the
same system studied in \cite{WeNew}.

A relevant aspect that arose from these studies is that there is a
kind of entrainment between the symmetry breaking of the NEP as
described in \cite{SSSR6}, together with a scaling of noise
intensity with system size as in \cite{SSSR4}.

In the first case we focused on a simple reaction-diffusion model
with a known form of the NEP \cite{WW,RL}, and --as in all three
cases-- considering the adiabatic limit and exploiting the two-state
approximation, we were able to clearly quantify the system size
dependence of the SNR. We have shown that in this case, SSSR is
associated to a NEP's symmetry breaking. For the second case we
analyzed the model of globally coupled nonlinear oscillators
discussed in \cite{SSSR4}, and have shown that it can also be
described within the NEP framework, but now SSSR arises through an
``effective" scaling of the noise intensity with the system size.
For the third studied case, we have obtained the exact form of the
noise-induced patterns (both the stable and unstable ones) as well
as the analytical expression of the NEP. The interplay of the
transition rates, that are essentially decreasing functions of
\(L\), and the behavior of $\Phi$, that increases with \(L\),
explain the existence of a maximum in the SNR for a specific length
of the system and a fixed noise intensity. What arose here, through
an alternative form of analysis, is that there is a kind of
entrainment between the symmetry breaking of the NEP as described in
\cite{SSSR6}, together with a scaling of noise intensity with the
system size as in \cite{SSSR4}.

The results found in this work clearly show that the
``nonequilibrium potential" (even if not known in detail, see for
instance \cite{quasi}) offers a very useful framework to analyze a
wide spectrum of characteristics associated to SR in spatially
extended or coupled systems. Within this framework the phenomenon of
SSSR looks, as other aspects of SR in extended systems
\cite{extend3a}, as a natural consequence of a breaking of the
symmetry of the NEP or to an effective scaling of the noise
intensity as in \cite{SSSR4}, or could be interpreted as an
entrainment between the two aspects.

In addition, in the first studied case, we have seen a new form of
resonant behavior through the variation of the coupling with the
surroundings. In such a case the system's response to an external
signal becomes more {\it robust}, that is {\it less sensitive} to
the precise value of the albedo parameter. This fact opens new
possibilities for analyzing and interpreting the behavior of some
biological systems \cite{neubio}.


As a final comment, the main difference between the first and third
cases when compared with the second one, is that in the two former
cases we have local interactions with albedo b.c. (covering the
range from Neumann to Dirichlet b.c.), while the latter has a non
local coupling together with boundary conditions that could be
assumed as Neumann. From our results, it is possible to argue that
the effective scaling of the noise that arises in the second case
comes from the non local interaction, and not from the b.c. To make
this aspect more obvious, we plan to study, within the present
framework, the competence between local and non-local spatial
couplings \cite{extend2b,extend3b}, that arise in some
multi-component models. Also, the consideration of more general
systems with several components will allow us to analyze the system
size dependence of SR between patterns in general
activator-inhibitor-like systems. All these aspects will be the
subject of further work.



\begin{acknowledgments}
The authors thanks A.D.S\'anchez and S.Mangioni for fruitful
discussions. HSW acknowledges the partial support from ANPCyT,
Argentine, and thanks to the European Commission for the award of
a {\it Marie Curie Chair} at the Universidad de Cantabria, Spain.
\end{acknowledgments}


\begin{thebibliography}{99}
\bibitem{foot1} Monostable means here that the function derive from
a monostable potential.
\end{thebibliography}
\end{document}